\documentclass[aps,pre,reprint,superscriptaddress,longbibliography,showpacs]{revtex4-2}

\usepackage{graphicx}
\usepackage{dcolumn}
\usepackage{bm}

\usepackage[figurename=Figure]{caption}
\usepackage{float}    
\usepackage{verbatim} 
\usepackage{amsmath}  
\usepackage{amssymb}  
\usepackage{listings} 
\usepackage{subfig}   
\usepackage{graphics}
\usepackage{latexsym,epsfig}

\usepackage[colorlinks=true,breaklinks=true,allcolors=blue]{hyperref}
\usepackage{lipsum}
\usepackage{dsfont}

\usepackage[scr=boondox]{mathalfa}

\def\Oo{\ensuremath{{\cal O}}} 
\def\Jj{\ensuremath{{\cal J}}} 

\def\Aa{\ensuremath{{\cal A}}}
 
\def\Hh{\ensuremath{{\cal H}}} 
\def\Nn{\ensuremath{{N}}}

\def\Cc{\ensuremath{{\cal C}}}
\def\Dd{\ensuremath{{\cal D}}}

\def\Tt{\ensuremath{{\cal T}}} 
\def\Gg{\ensuremath{{\cal G}}} 
\def\Zz{\ensuremath{{\cal Z}}}

\def\Qq{\ensuremath{{\cal Q}}}

\def\Lll{\ensuremath{\mathscr{L}}}
\usepackage[scr=boondox]{mathalfa}

\def\RR{\ensuremath{{\mathbb{R}}}}
\def\p{\ensuremath{{\partial}}}
\def\i{\ensuremath{{\imath}}}
\def\iu{\ensuremath{{\imath}}}
\def\Im{\ensuremath{{\operatorname{Im}}}}
\def\Re{\ensuremath{{\operatorname{Re}}}}

\usepackage{tikz,bm} 
\usepackage{circuitikz}

\definecolor{mycolor}{rgb}{1,0.2,0.3}
\definecolor{brightgreen}{rgb}{0.4, 1.0, 0.0}
\definecolor{britishracinggreen}{rgb}{0.0, 0.26, 0.15}
\definecolor{cadmiumgreen}{rgb}{0.0, 0.42, 0.24}
\definecolor{ceruleanblue}{rgb}{0.16, 0.32, 0.75}
\definecolor{darkelectricblue}{rgb}{0.33, 0.41, 0.47}
\definecolor{darkpowderblue}{rgb}{0.0, 0.2, 0.6}
\definecolor{darktangerine}{rgb}{1.0, 0.66, 0.07}
\definecolor{emerald}{rgb}{0.31, 0.78, 0.47}
\definecolor{palatinatepurple}{rgb}{0.41, 0.16, 0.38}
\definecolor{pastelviolet}{rgb}{0.8, 0.6, 0.79}

\begin{document}
	
	\preprint{APS/123-QED}
	
	\title{Dynamics and Charge Fluctuations in Large-q Sachdev-Ye-Kitaev Lattices}
	
	\author{Rishabh Jha}
	\email{rishabh.jha@uni-goettingen.de}
	\author{Jan C. Louw}
	\email{jan.louw@theorie.physik.uni-goettingen.de}
	\affiliation{%
		Institute for Theoretical Physics, Georg-August-Universit\"{a}t G\"{o}ttingen, Germany
	}%

	
	
	
	
	\begin{abstract}
		It is known that the large-$q$ complex Sachdev-Ye-Kitaev (SYK) dot thermalizes instantaneously under rather general dynamical protocols. We consider a lattice of such dots coupled together, allowing for $r/2$ body hopping of particles between nearest neighbors. We develop a rather general analytical framework to study the dynamics to leading order in $1/q$ on such a lattice, allowing for arbitrary time-dependent couplings, hence general dynamical protocols. We find that the physics of the diffusive case $r>2$ is effectively the same as the kinetic case $r=2$, assuming $r=\Oo(q^0)$. Remarkably, we find that the local charge densities $\Qq_i$ form a closed set of equations. They however only show fluctuations of the order $\Oo(\Qq_i/q)$, hence remaining constant in the limit $q\rightarrow \infty$. Despite this effective lack of charge dynamics, the dots do not in fact behave as isolated lattice sites which would thermalize instantaneously. Indeed, we show via a proof by contradiction that such instantaneously thermalize is not generally possible for a connected lattice. Importantly, the results are shown to be independent of the dimensionality of the lattice. 
	\end{abstract}
	
	\maketitle
	

	\section{Introduction}
	\label{1}
	
	The Sachdev-Ye-Kitaev (SYK) model is a generalization of the Sachdev-Ye model \cite{Sachdev1993} proposed by Kitaev \cite{Kitaev2015} as a model for quantum holography where $q$ Majorana fermions interact via random matrix coupling in a total of $\Nn$ particles. The SYK model is a $0+1$ dimensional strongly coupled quantum field theory. Given this, it has attracted attention due to its analytical tractability in the large-$\Nn$ limit where the Schwinger-Dyson equations can be written in a closed form. This is despite the model being maximally chaotic \cite{Maldacena2016a}. 
	
	To bring the model closer in contact with a condensed matter system, one usually considers complex charged fermions \cite{Gu2020,Song2017,Fu2018}. Such a natural generalization is known as the complex SYK model. Unlike the Majorana case, here the number of particles is a definable quantity associated with a conserved $\text{U}(1)$ charge due to the presence of fermionic charges. This charge may be varied by introducing a chemical/mass potential term in the Hamiltonian. When considering this model at charge neutrality, the Majorana case is recovered. 
	
	Despite its simplification at large $N$, the model is usually only fully solvable via numerics. At low energies, an emergent conformal symmetry does however allow one to extract certain analytical results \cite{Fu2018}. In considering $q$-body interactions, one may in fact analytically solve the model order by order in $1/q$ \cite{Tarnopolsky2019}. The leading order results are often qualitatively reflective of the $q\ge 4$ models. For instance, quantitatively and qualitatively similar phase transitions are observed at all $q\ge 4$ \cite{Louw2023,Ferrari2019,Azeyanagi2018,Sorokhaibam2020}. This system has a tendency to thermalize rapidly \cite{Eberlein2017}. In particular, given a general non-equilibrium protocol to a single large-$q$ SYK model, the system will thermalize instantaneously \cite{Louw2022}. A better understanding of this thermalization process is still lacking. For instance, under which conditions a large-$q$ SYK model would not thermalize instantaneously? 
	
	Moreover, one may for instance study charge transport along a chain of complex SYK dots. In this setup each lattice site (or blob for a better physical picture) is occupied by a complex SYK model. The blobs are connected by transport terms with nearest neighbor hopping. Such a one-dimensional chain is intimately connected to strongly correlated quantum matter and strange metals, which are considered to be at the heart of modern condensed matter theory. They have been shown to exhibit non-Fermi liquid transport behavior \cite{Chowdhury2018, Chowdhury2020,Song2017}, for instance, a linear in $T$ resistivity \cite{Song2017}. In other words, their behavior is not captured by a quasi-particle picture.  
	
	One analytically tractable construction considers such a chain where each lattice site is occupied by a large-$q$ complex SYK model. Naturally, this construction has been studied in the literature \cite{Zanoci2022}, where the transports terms also include $q/2$-body hopping. The analytically tractable property of the large-$q$ SYK model is ten exploited to extract exact analytical results. This then provides analytical insight into strongly correlated matter. For instance, by imposing uniform temperature and chemical potential gradients, thermoelectric transport properties may be calculated. 
	
	In this work, we consider a similar construction: a one-dimensional lattice where each blob has a large-$q$ complex SYK model and the blobs are connected by $r/2$-particle transport between nearest neighbor. Standard (quadratic) hopping would correspond to $r=2$, while we also allow for diffusive hopping $r>2$. We consider $r$ to be order of $\Oo(q^0)$. Due to analytic tractability, this has become one of the prototypical examples for analytic calculations of various transport properties. We develop a rather general framework required to study the dynamical properties of this system. Our framework is well suited to handle general dynamical protocols such as quenches and ramps in order to study the non-equilibrium behavior of the system. Such dynamical protocols will be the focus of this work, instead of temperature and chemical potential gradients considered in \cite{Zanoci2022}. The $1/q$ expansion drastically simplifies the analysis. For instance, we find that to leading order in $1/q$, the equation of motion for the charge is closed under the charge density. In other words, the complicated Green's functions do not enter. 
	
	With this, we may analytically calculate the charge transport dynamics in the system. In particular, we focus on a quench from a disconnected, with transport terms switched off, to a connected chain. We find a discrete wave equation for the charge transport. Solving this equation, we show how current flows directly after transport is switched on. We find however that, in the large $q$ limit, the local charge remains constant. From this, one might assume that each dot behaves as an isolated (instant thermalizing) large $q$ SYK system. We show that this is in fact not the case. This is done via a proof by contradiction. Assuming the chain does thermalize instantaneously, implies a certain consistency relation. This relation is not satisfied for our quench, hence the system cannot thermalize instantaneously. One may, however, consider when the consistency relation would be fulfilled. This would then provide a set of conditions under which instantaneous thermalization cannot be ruled out. One of these cases is when all transport coefficients are set to zero. Thus, our proof is consistent with the instantaneous thermalization of isolated blobs of large-$q$ complex SYK models \cite{Louw2022}. 
	
	Lastly, we show that these results are immediately generalizable to a $d$-dimensional lattice.

	\section{Model \& Framework}
	\label{2}
	
	\subsection{Model}
	\label{2.1}
\begin{figure}
		\includegraphics[width=0.99\linewidth]{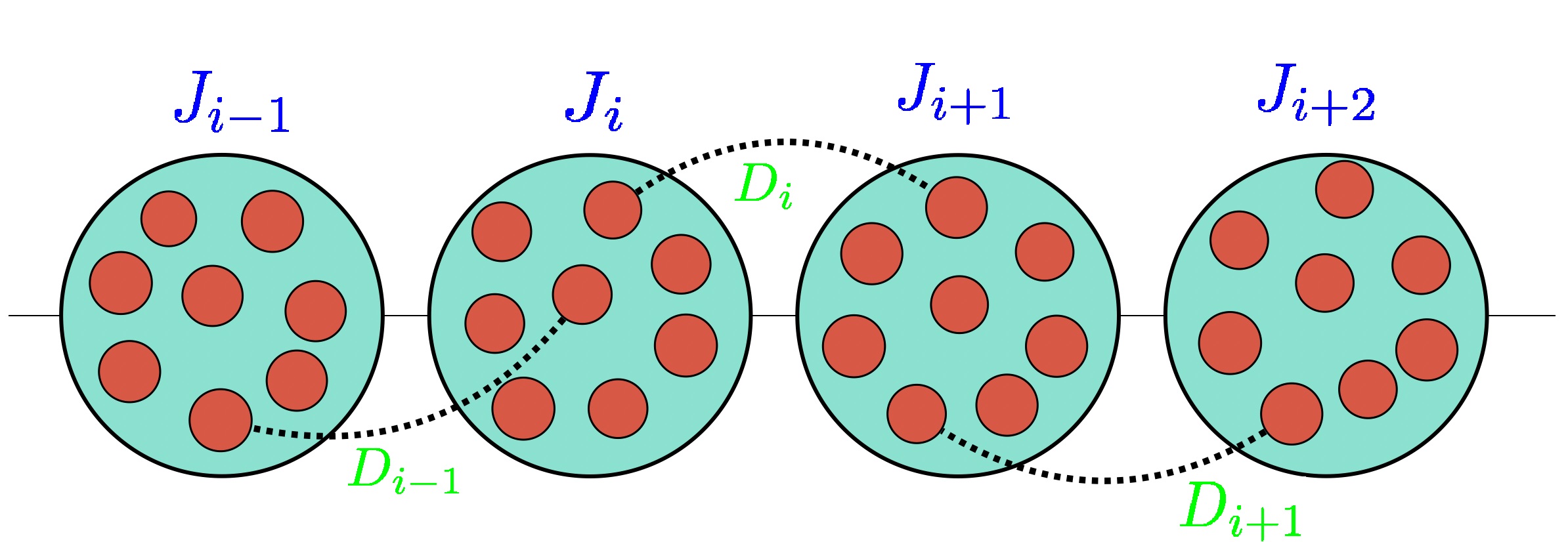}
		\caption{One-dimensional chain where each blob contains large-$q$ complex SYK model with on-site strength for $i^{\text{th}}$ site is given by $J_i$ while the nearest neighbor coupling strength is given by $D_i$.}.
		\label{fig}
	\end{figure}

	We consider a chain consisting of $2L$ lattice blobs where each blob is occupied by a large-$q$ complex SYK model. The Hamiltonian is given as follows (see Fig. \ref{fig}):
	\begin{equation}
		\Hh(t) = \sum_{i=1}^{2L}  \left(\Hh_{i}(t) + \Hh_{i \to i+1}(t) + \Hh_{i \to i+1}^\dag(t)\right) 
		\label{HamChain}
	\end{equation}
	where the on-site large-$q$ complex SYK Hamiltonian is given by
	\begin{equation}
		\begin{aligned}
			\Hh_{i}(t)
			&= J_i(t) \hspace{-1mm} \sum\limits_{\substack{ \{\bm{\mu}\}_1^{q/2} \\ \{\bm{\nu}\}_1^{q/2} }} \hspace{-2mm}X(i)^{\bm{\mu}}_{\bm{\nu}} c^{\dag}_{i;\mu_1} \cdots c^{\dag}_{i; \mu_{q/2}} c_{i;\nu_{q/2}}^{\vphantom{\dag}} \cdots c_{i;\nu_1}^{\vphantom{\dag}}
		\end{aligned}
		\label{hi}
	\end{equation}
	summing over $\{\bm{\nu}\}_{1}^{q/2} \equiv 1\le \nu_1<\cdots< \nu_{q/2}\le\Nn$. The transport of $r/2$ fermions from site $i$ to $i+1$ is given by
	\begin{equation}
		\begin{aligned}
			\Hh_{i \rightarrow i+1}(t) 
			&= D_i(t) \hspace{-1mm} \sum\limits_{\substack{ \{\bm{\mu}\}_1^{r/2} \\ \{\bm{\nu}\}_1^{r/2} }} \hspace{-2mm}Y(i)^{\bm{\mu}}_{\bm{\nu}} c^{\dag}_{i+1;\mu_1} \cdots c^{\dag}_{i+1; \mu_{\frac{r}{2}}} c_{i;\nu_{\frac{r}{2}}}^{\vphantom{\dag}} \cdots c_{i;\nu_1}^{\vphantom{\dag}}.
		\end{aligned}
		\label{hi to i+1}
	\end{equation}
	The operators $c^\dag_{i;\alpha}$ and $c_{i;\alpha}$ are spinless fermionic creation and annihilation operators (associated with lattice site $i$ and flavor $\alpha$)  respectively. Here $J_i(t)$ and $D_i(t)$ are the coupling strengths of the on-site and the transport interactions respectively. Both $X(i)^{\bm{\mu}}_{\bm{\nu}}$ and $Y(i)^{\bm{\mu}}_{\bm{\nu}}$ are independent random matrices whose components are derived from Gaussian ensemble with zero mean and variances
	\begin{equation}
		\begin{aligned}
			\overline{|X|^2} &= \frac{q^{-2}((q / 2) !)^2}{(N / 2)^{q-1}} \\
			\overline{|Y|^2} &= \frac{1}{q}\frac{(1 / r)((r / 2) !)^2}{(N / 2)^{r-1}}.
		\end{aligned}
	\end{equation}
	In order to introduce competition between the transport terms and the on-site interactions, we need to introduce a $1/q$ scaling in the variance for the random matrix $Y(i)^{\bm{\mu}}_{\bm{\nu}}$.
	For $r=2$, the hopping is kinetic while $r>2$ corresponds to a \emph{diffusive}-type transport. We also allow for a local mass term of the form
	\begin{equation}
		\Hh_0(t) = - \sum_i \dot{\eta}_i(t) N\Qq_i, \quad  \Qq_{i} \equiv \frac{1}{N}\sum_{\alpha=1}^N[c_{i;\alpha}^\dag c_{i;\alpha}-1/2]
		\label{mass}
	\end{equation} 
	where $\Qq_{i}$ is the local charge density on the $i^{\text{th}}$ blob, $\eta_i(t)$ is an arbitrary function playing a role as of chemical potential and $\Nn$ is the number of particles on the blob. Using the benefit of hindsight, we have introduced the derivative $\dot{\eta}_i (t)$ here. With this, the total Hamiltonian would be $\Hh(t)+\Hh_0(t)$. Although the Hamiltonian is of diffusive type, it is an interesting feature of charged SYK lattices that their transport properties are more dependent on the ratio between $r$ and $q$. For instance, one may show that, beyond the coherent regime, that the resistivity behaves as $\rho \sim T^{2(r/q-1)}$ \cite{Cha2020}. In other words, for large $q$, the properties will not be strongly affected by the order of the transport. To have qualitative and quantitative diffusive transport, in the large $q$ limit, one would thus have to scale $r$ with $q$.
	
	We can assume both periodic as well as non-periodic boundary conditions over the lattice. If we assume periodic boundary conditions, then we consider the blob $2L+1 \equiv_{2L} 1$ where the subscript $2L$ denotes the periodicity. For non-periodic boundary conditions, we set the coupling strength transporting fermions from blobs $2L \rightarrow 2L+1$ equal to zero.

	\subsection{Schwinger-Dyson Equations}
	\label{2.2}
	Our main interest is in the non-equilibrium dynamics of our chain. We consider the general time evolution along the Keldysh contour $\Cc$, with a focus on the flavor averaged Green's functions which are defined as follows:
	\begin{equation}
		\Gg_{ij}(t_1,t_2) \equiv \frac{-1}{N} \sum\limits_{\alpha=1}^{N}\langle \Tt_{\Cc} c_{i;\alpha}(t_1) c_{j;\alpha}^{\dag}(t_2) \rangle.
	\end{equation}
	Here $\Tt_\Cc$ is the time ordering operator w.r.t. $\Cc$. We are in the Heisenberg picture in the Keldysh formalism where the averaging is taken with respect to the initial non-interacting Hamiltonian \cite{Kamenev2009}. These Green's functions are the solution to Dyson's equation $\hat{\Gg}  = \hat{\Gg}_0+\hat{\Gg} _0\hat{\Sigma} \hat{\Gg} = \hat{\Gg}_0+  \hat{\Gg} \hat{\Sigma} \hat{\Gg}_0$. Here $\hat{\Sigma}$ is the $2L$ by $2L$ self-energy matrix, which is diagonal $\hat{\Sigma}_{i,i+d} \equiv \delta_{d0} \Sigma_{i}$ for the SYK chain. We consider diagonal initial conditions $\hat{\Gg}_{i,i+d} = \delta_{d0} \Gg_{ii}$, hence we need to only consider the local Green's functions $\Gg_{i} \equiv \Gg_{ii}$ corresponding to each blob in the lattice. 
	
	Similar to one complex SYK model, we write down the partition function corresponding to the Hamiltonian in eq. (\ref{HamChain}), introduce the fields $\Gg_i$ and $\Sigma_i$ through delta functions and then integrate out the fermions to get
	\begin{equation}
		\mathcal{Z} = \int \mathcal{D}\Gg_i \mathcal{D}\Sigma_i e^{-\Nn S_{0;i}[\Gg, \Sigma] - \Nn S_{I;i}[\Gg, \Sigma]}
	\end{equation} 
	where $S_{0;i}$ is the effective action on $i^{\text{th}}$ blob while $S_{I;i}$ is the effective transport action corresponding to the transport from blobs $i-1$ and $i+1$ to and from the $i^{\text{th}}$ blob. They are given by \footnote{A combinatorial argument to get the effective action (consequently the Schwinger-Dyson equations) is provided in Appendix \ref{appendix A}.}
	\begin{widetext}
		\begin{equation}
			\begin{aligned}
				S_{0;i} = - \operatorname{Tr} \ln\left(\Gg_{o;i}^{-1} - \Sigma_i\right) - \int d t_1 d t_2\left(\Sigma_i\left(t_1, t_2\right) \Gg_i\left(t_2, t_1\right)-\frac{J_i(t_1)J_i(t_2)}{2q^2}\left(-4\Gg_i\left(t_1, t_2\right) \Gg_i\left(t_2, t_1\right)\right)^{\frac{q}{2}}\right)
			\end{aligned}
			\label{noninteracting action}
		\end{equation}
	\end{widetext} 
	as well as
	\begin{equation}
		S_{I;i} \equiv \int dt_1 dt_2 \hspace{1mm}  \Lll_{I;i}[\Gg](t_1,t_2) 
		\label{interacting action}
	\end{equation}
	where the Lagrangian $\Lll_{I;i}$ corresponding to the Hamiltonian (eq. (\ref{HamChain})) is given by
	\begin{equation}
		\begin{aligned}
			\Lll_{I;i}[\Gg]=&  \frac{D_{i-1}^*(t_1) D_{i-1}(t_2)}{q r} [-4 \Gg_{i-1}(t_1,t_2) \Gg_{i}(t_2,t_1)]^{r/2} \\
			&+  \frac{D_{i}(t_1)D_{i}^*(t_2)}{qr} [-4\Gg_{i+1}(t_1,t_2) \Gg_{i}(t_2,t_1)]^{r/2}.
			\label{interacting lagrangian}
		\end{aligned}
	\end{equation}
	This is similar to that found in \cite{Zanoci2022}. The differences lie only in a redefinition of the couplings and allowing them to be time-dependent. Clearly when $D_i = 0$ $\forall$ $i$, we obtain the disconnected SYK blobs whose effective action is given by eq. (\ref{noninteracting action}) as expected. Therefore, the total effective action is given by
	\begin{equation}
		S_{\text{eff},i} = S_{0;i} + S_{I;i}
		\label{action}
	\end{equation}
	where $S_{0;i}$ and $S_{I;i}$ are given in eq. (\ref{noninteracting action}) and eq. (\ref{interacting action}, \ref{interacting lagrangian}) respectively.
	
	Having obtained the effective action in the large-$\Nn$ limit for the Hamiltonian (eq. (\ref{HamChain})), we take its functional derivative to get the local self-energy $\Sigma_i$. We see that there are two contributions to the local self-energy $\Sigma_i$, namely $ \Sigma_{J,i}$ and $\Sigma_{D,i}$ where $\Sigma_{J,i}$ is the on-site contribution and $\Sigma_{D,i}$ is the transport contribution at the $i^{\text{th}}$ blob. Thus we can write
	\begin{equation}
		\Sigma_i(t_1, t_2)=  \Sigma_{J,i} (t_1, t_2)+ \Sigma_{D,i}(t_1, t_2)
		\label{se1}
	\end{equation}
	where	
	\begin{equation}
		\scalebox{0.95}[1]{$q\Sigma_{J, i} =  2J_{i}(t_1)J_{i}(t_2)\Gg_{i}(t_1,t_2) \left[-4\Gg_{i}(t_1,t_2) \Gg_{i}(t_2,t_1)\right]^{\frac{q}{2}-1}$} 
		\label{se2}
	\end{equation}
	and
	\begin{widetext}
		\begin{equation}
			\begin{aligned}
				q\Sigma_{D, i} =  D_{i-1}^*(t_1)D_{i-1}(t_2) [-4\Gg_{i}(t_2,t_1) & \Gg_{i-1}(t_1,t_2)]^{\frac{r}{2}-1} 2\Gg_{i-1}(t_1,t_2) \\
				&+  D_{i}(t_1)D_{i}^*(t_2) [-4\Gg_{i}(t_2,t_1)\Gg_{i+1}(t_1,t_2)]^{\frac{r}{2}-1} 2\Gg_{i+1}(t_1,t_2)
			\end{aligned}
			\label{se3}
		\end{equation}
	\end{widetext}
	from which one can read off the same conjugate relation $\Sigma_i(t_1, t_2)^\star = \Sigma_i(t_2, t_1)$ as the Green's functions.

	\subsection{Kadanoff-Baym Equations}
	\label{2.3}
	Using Langreth rules \cite{Stefanucci2010}, Dyson's equations yield the Kadanoff-Baym (KB) equations which can be expressed as follows where we take $t_2>_{\Cc}t_1$ without loss of generality (chosen to lie on different halves of $\Cc$) \cite{Louw2022}:
	\begin{equation}
		\begin{aligned}
			\left[\partial_{t_1} - \right. & \left. \iu \dot{\eta}_i(t_1) \right]\Gg_i^{\gtrless}(t_1,t_2) \\
			&= \int_{t_1}^{t_2} dt_3 \hspace{1mm} \left( \Sigma_i^{\gtrless}(t_1,t_3) \Gg_i^{A}(t_3,t_2) \right) - \frac{\iu}{2 q} \alpha_i(t_1,t_2).
		\end{aligned}
		\label{kb}
	\end{equation}
	Here the forward/backward Green's functions $\Gg_i^{\gtrless}$ correspond to $t_1$ being ahead/behind of $t_2$ on the Keldysh contour $\Cc$ respectively. They are defined as
	\begin{equation}
		\begin{aligned}
			& \mathcal{G}_i^{>}\left(t_1, t_2\right) \equiv-\frac{1}{N} \sum_{\mu=1}^N\left\langle c_{i;\mu}\left(t_1\right) c_{i;\mu}^{\dagger}\left(t_2\right)\right\rangle \\
			& \mathcal{G}_i^{<}\left(t_1, t_2\right) \equiv \frac{1}{N} \sum_{\mu=1}^N\left\langle c_{i;\mu}^{\dagger}\left(t_2\right) c_{i;\mu}\left(t_1\right)\right\rangle.
		\end{aligned}
		\label{gtrless gf}
	\end{equation}
	
	For $t_2 >_{\Cc} t_1$ the advanced Green's function is given by
	\begin{equation}
		\Gg_i^A(t_1, t_2) = \Gg_i^<(t_1, t_2) - \Gg_i^>(t_1, t_2), \label{advanced g}
	\end{equation}
	while it is zero otherwise. Finally, $\alpha_i$ in eq. (\ref{kb}) is defined as
	\begin{equation}
		\begin{aligned}
			\alpha_i(t_1,t_2) \equiv & \hspace{1mm}\iu \int_{t_0}^{t_1} dt_3 \hspace{1mm} q\Sigma_i^{>}(t_1,t_3) 2\Gg_i^{<}(t_3,t_2) \notag\\
			&  -\iu \int_{t_0-\iu\beta}^{t_1} dt_3 \hspace{1mm} q\Sigma_i^{<}(t_1,t_3)2\Gg_i^{>}(t_3,t_2) \label{alpha}
		\end{aligned}
	\end{equation}
	where we define the forward and backward self-energies $\Sigma_i^{\gtrless}$ in the same manner as the Green's functions $\Gg_i^{\gtrless}$ whose explicit expressions can be obtained using eqs. (\ref{se1}, \ref{se2}, \ref{se3}) as follows:
	\begin{widetext}
		\begin{equation}
			\begin{aligned}
				q\Sigma_{i}^>(t_1,t_2) =& \hspace{1mm}2 J_{i}(t_1)J_{i}(t_2) [-4\Gg_{i}^>(t_1,t_2)  \Gg_{i}^<(t_2,t_1)]^{q/2-1} \Gg_{i}^>(t_1,t_2) \\
				&+  2D_{i-1}(t_2) D_{i-1}^*(t_1) [-4\Gg_{i-1}^>(t_1,t_2) \Gg_{i}^<(t_2,t_1)]^{r/2-1} \Gg_{i-1}^>(t_1,t_2) \\
				&+  2D_{i}(t_1)D_{i}^*(t_2) [-4\Gg_{i+1}^>(t_1,t_2)\Gg_{i}^<(t_2,t_1)]^{r/2-1} \Gg_{i+1}^>(t_1,t_2) \\
				q\Sigma_{i}^<(t_2,t_1) =&\hspace{1mm} 2 J_{i}(t_1)J_{i}(t_2)  [-4\Gg_{i}^>(t_1,t_2)   \Gg_{i}^<(t_2,t_1) ]^{q/2-1} \Gg_{i}^<(t_2,t_1) \\
				&+  2D_{i-1}(t_1)D_{i-1}^*(t_2) [-4\Gg_{i}^>(t_1,t_2) \Gg_{i-1}^<(t_2,t_1)]^{r/2-1} \Gg_{i-1}^<(t_2,t_1) \\
				&+  2D_{i}(t_2)D_{i}^*(t_1) [-4\Gg_{i}^>(t_1,t_2)\Gg_{i+1}^<(t_2,t_1)]^{r/2-1} \Gg_{i+1}^<(t_2,t_1)
			\end{aligned}
			\label{se gtrless}
		\end{equation}
	\end{widetext}
	from them one can see $\Sigma_i^{\gtrless}(t_1, t_2)^\star = \Sigma_i^{\gtrless}(t_2, t_1)$. The first term in both the expressions are the on-site contributions to the self-energies while the second and the third terms are the transport contributions. Also for the kinetic case where $r=2$, these expressions considerably simplify, although we will always consider a general $r$ $\left( = \Oo(q^0) \right)$ in this work.
	We assume the weakening of initial conditions at initial time $t_0 \rightarrow -\infty$ (Bogoliubov principle) \cite{Semkat1999, Pourfath2007}. Under this assumption, the imaginary part of the contour in $\alpha_i$ in eq. (\ref{alpha}) is ignored.
	Considering equal times ($t_2 \rightarrow t_1$), the KB equations in eq. (\ref{kb}) reduce to 
	\begin{equation}
		\alpha_i(t_1,t_1) = 2 \iu q [\partial_{t_1}-\iu\dot{\eta}_i(t_1)]\Gg^{<}_{i}(t_1,t_1^+)
		\label{equal time kb}
	\end{equation}
	where the limit $t_1^+ \to t_1$ is taken only after differentiating.

	\subsection{Expectation Values of Energy}
	\label{2.4}
	We are interested in finding the expectation values of local on-site energy as well as the transport energy. Considering the explicit definition of the backward Green's function, we get for the right-hand side of eq. (\ref{equal time kb})
	\begin{equation}
		2\iu q[\partial_{t_1}-\iu\dot{\eta}_i(t_1)]\Gg^{<}_{i}(t_1,t_1^+) = \frac{2q}{N}\langle \sum\limits_{\alpha} c_{i,\alpha}^\dagger \left[ c_{i,\alpha},\Hh \right] \rangle (t_1).
		\label{rhs}
	\end{equation} 
	We can further evaluate using the identity $\left[ c,\Hh \right] = \partial_{c^\dagger} \Hh$ where $\partial_{c^\dagger}$ anti-commutes with fermionic operators. This leads to
	\begin{equation}
		2\sum_{\alpha}  c_{i,\alpha}^\dag [c_{i,\alpha},\Hh] =  q\Hh_i + r\Hh_{i\to i+1}^\dag + r \Hh_{i-1\to i}
		\label{gm rule}
	\end{equation}
	where we used the identity for any even $n-$body interaction term and a generalized Galitskii-Migdal sum rule \cite{Louw2022,Stefanucci2010} ($n$ can be $q$ or $r$ in our case depending on whether we are dealing with the on-site Hamiltonian or the transport Hamiltonian respectively) 
	Plugging eq. (\ref{gm rule}) in eq. (\ref{rhs}), we get
	\begin{equation}
		\begin{aligned}
			2\iu q[&\partial_{t_1}-\iu\dot{\eta}_i(t_1)]\Gg^{<}_{i}(t_1,t_1^+) \\
			&= \frac{q^2}{\Nn} \langle \Hh_i \rangle (t_1) + \frac{qr}{\Nn} \langle \Hh_{i \rightarrow i+1}^\dagger \rangle(t_1) + \frac{qr}{\Nn} \langle \Hh_{i-1 \rightarrow i} \rangle (t_1).
		\end{aligned}
	\end{equation}
	According to the equal-time KB equations (eq. (\ref{equal time kb})), we know that this expression is equal to $\alpha_i(t,t)$. Accordingly, we define the local and transport expectation values as
	\begin{equation}
		\epsilon_{i}(t_1) \equiv  \frac{q^2}{\Nn} \langle \Hh_i\rangle(t_1), \quad \epsilon_{i\to i+1}(t_1) \equiv \frac{q^2}{\Nn} \langle \Hh_{i\to i+1} \rangle(t_1).
		\label{ener}
	\end{equation}
	Thus the equal-time KB equations become
	\begin{equation}
		\alpha_i(t_1,t_1) = \epsilon_i(t_1)+\frac{r}{q} \left[ \epsilon_{i\to i+1}^*(t_1)+\epsilon_{i-1\to i}(t_1) \right].
		\label{kb energy}
	\end{equation}
	We can extract the correspondence between these expectation values and integrals of the Green's functions and self-energies by taking derivatives of the total effective action (eqs. (\ref{noninteracting action}, \ref{interacting action}, \ref{action})) with respect to the corresponding coupling constants ($J_i$ or $D_i$). This leads to the following expressions for the on-site and transport contributions
	\begin{widetext}
		\begin{equation}
			\begin{aligned}
				\epsilon_i(t_1)  &= \Im\int_{-\infty}^{t_1}dt_2 \hspace{1mm} 2 J_i(t_1) J_{i}(t_2) [-4\Gg_{i}^<(t_1,t_2) \Gg_{i}^>(t_2,t_1)]^{q/2}  \\
				\epsilon_{i \rightarrow i+1}(t_1) &= \int_{-\infty}^{t_1}d t_2 \hspace{1mm} \frac{\iu q}{r} D_{i}^*(t_1) D_{i}(t_2)  \left[ \left(-4\Gg_{i+1}^>(t_1,t_2)\Gg_{i}^<(t_2,t_1)\right)^{r/2}-\left(-4\Gg_{i}^>(t_2,t_1) \Gg_{i+1}^<(t_1,t_2)\right)^{r/2} \right] 
			\end{aligned}
			\label{full energy}
		\end{equation}
	\end{widetext}
	respectively, where $\Im$ denotes the imaginary part. This can be verified by plugging these expressions in eq. (\ref{kb energy}) and using the definition of $\alpha_i$ in eq. (\ref{alpha}) for $t_1 = t_2$. 
	
	\subsection{Functional Form of Green's Functions}
	\label{2.5}
	We express our Green's functions in the following form 
	\begin{equation}
		\Gg_{i}^{\gtrless}(t_1,t_2)  = \mp \left( \frac{1}{2} \mp \Qq_i(t) \right) e^{\i\eta_{i}(t_1,t_2) + g_{i}^\gtrless(t_1,t_2)/q} 
		\label{large q gf}
	\end{equation} 
	where we have defined the time average $t \equiv (t_1 + t_2)/2$ \footnote{Sometimes this is also denoted as $t_{12}^{+}$.} and $\Qq$ is defined in eq. (\ref{mass}). Considering the definitions of Green's functions in eq. (\ref{gtrless gf}) and the local charge density, we have $g_{i}^{\gtrless}(t,t) =0$ \footnote{In other words, for $t_1 = t_2 + \epsilon$ and $t_2 = t$, for small $\epsilon$ we can write $	\Gg_{i}^{\gtrless}(t+\epsilon,t) \equiv \Qq_{i}(t) -\text{sgn}(\epsilon)/2$.}. Similar to the Green's functions, for $t_1 >_{\Cc} t_2$ we have $g_{i}(t_1,t_2) = g_{i}^>(t_1,t_2) $ while for $t_1 <_{\Cc} t_2$ we have $g_{i}(t_1,t_2) = g_{i}^<(t_1,t_2) $. Given the expression \eqref{large q gf}, the proof is shown in \cite{Louw2023} that $g_{i}^\gtrless = \Oo(q^0)$, implying that it is a good starting point of a $1/q$ expansion. The exponential form also yields a larger overlap with the exact $q=4$ solution \cite{Tarnopolsky2019}. In the interaction picture, we have equations of motion such as $\dot{c}(t) = \iu [-\dot{\eta}(t) c^\dag c, c](t) = \iu \dot{\eta}(t) c(t)$ solved by $c(t) = e^{\iu  \eta(t)} c$ and similarly $c^\dag(t) = e^{-\iu \eta(t)} c^\dag$. These suggest to conveniently define the following quantity:
	\begin{equation}
		\eta_i(t_1, t_2) \equiv \eta_i(t_1)-\eta_i(t_2) 
		\label{g i eta}
	\end{equation}
	where the KMS relation for $\Gg_i^{\gtrless}$ provides leading order scaling in $q$ as $\eta_i = \Oo(\Qq) = \Oo(q^{-1/2})$ \cite{Sorokhaibam2020}.
	
	As shown in \cite{Louw2022}, the on-site energy density is bounded as $\left| \epsilon_i \right| \leq 2 e^{-q \Qq^2} J_i$. Thus for nontrivial interactions, we focus on small charge densities in the large-$q$ limit as $\Qq_i = \Oo(q^{-1/2})$. Accordingly, we can conveniently move the charge densities appearing in eq. (\ref{large q gf}) into the exponential as
	\begin{equation}
		1 \mp 2\Qq_i(t) \sim e^{-2\Qq_i(t)^2\mp 2 \Qq_i(t)} 
	\end{equation} 	
	which is correct to quadratic order in charge density. Therefore, plugging this in eq. (\ref{large q gf}), we explicitly have for the Green's functions at leading order in $1/q$
	\begin{equation}
		\begin{aligned}
			-2\Gg_{i}^>(t_1,t_2)  &= e^{-2[\Qq_i(t)+\Qq_i(t)^2] + \i\eta_{i}(t_1,t_2) + g_{i}^>(t_1,t_2)/q} \\
			2\Gg_{i}^<(t_1,t_2) &= e^{-2 [-\Qq_{i}(t) + \Qq_{i}(t)^2] +\i\eta_{i}(t_1,t_2) + g_{i}^<(t_1,t_2)/q}.
		\end{aligned}
		\label{functional gf}
	\end{equation}
	We can use these leading order in $1/q$ results to get the explicit expressions for self-energies $\Sigma_i^{\gtrless}(t_1, t_2)$ using eq. (\ref{se gtrless}). The final expressions are quite lengthy but straightforward to obtain. We present the following results for the kinetic hopping case where $r=2$ to leading order in $1/q$ in the large-$q$ limit:
	\begin{widetext}
		\begin{equation}
			\begin{aligned}
				-q\Sigma_{i}^{>}(t_1,t_2) &= J_{i}(t_1)J_{i}(t_2)e^{-2q \Qq_{i}(t)^2} e^{\left( g_i^>(t_1,t_2) + g_i^<(t_2,t_1) \right)/2} \\
				q\Sigma_{i}^{<}(t_2,t_1) &= J_{i}(t_1)J_{i}(t_2)e^{-2q \Qq_{i}(t)^2} e^{\left( g_i^>(t_1,t_2) + g_i^<(t_2,t_1) \right)/2}.
			\end{aligned}
			\label{functional se}
		\end{equation}
	\end{widetext}
	Thus we obtained the functional form for the large-$q$ expansion of Green's functions in eq. (\ref{functional gf}) which also led us to the functional form for the large-$q$ expansion of self-energies where we presented the results for the kinetic hopping case in eq. (\ref{functional se}). We already know that $g_i^{\gtrless}(t_1, t_2) = \Oo(q^0)$ and $\eta_i = \Oo(q^{-1/2})$. Then starting from eq. (\ref{kb}), we use these results to obtain the leading order KB in $q$ in Sec. \ref{4.1}.
	
	\section{Charge Transport}
	\label{3}
	We are interested in studying the non-equilibrium charge transport dynamics in the chain where there is a quench done at $t=0$. Before we deal with the quench dynamics, we develop a general formalism to study the charge transport. Using the functional form of the Green's functions in eq. (\ref{large q gf}) where we know already that $g_{i}^{\gtrless} (t_1,t_1) = 0$, we have that $\Gg^<_i(t_1,t_1) = \Qq_i(t_1)+1/2$ implies that
	\begin{equation}
		\begin{aligned}
			\dot{\Qq}_{i}(t_1) = \p_{t}\Gg_{i}^<(t_1,t_1^+) + \partial_{t_1}\Gg_{i}^<(t_1^+,t_1)
		\end{aligned}
	\end{equation}
	where the limit $t_1^+ \rightarrow t_1$ is taken only after the derivative has been taken. Due to the structure of the right-hand side, we are interested in the real part of the KB equation at equal time in eq. (\ref{kb}). We note that the mass term $\dot{\eta}_i$ is real, the real part of eq. (\ref{kb}) at equal time takes the following form in terms of the change in local charge density
	\begin{equation}
		\dot{\Qq}_{i}(t_1) = \Im \left[ \alpha_i(t_1,t_1) \right]/q .
		\label{Qdot}
	\end{equation}
	But we already know the form of $\alpha_i$ from eq. (\ref{kb energy}) where using the explicit form of $\epsilon_{i \rightarrow i+1}(t_1)$ from eq. (\ref{full energy}), we get that $\epsilon_{i \rightarrow i+1}(t_1)^\star = - \epsilon_{i \rightarrow i+1}(t_1)$. This then yields
	\begin{equation}
		\dot{\Qq}_{i}(t_1) = \frac{r}{q^2} \Im \left[ \epsilon_{i-1\to i}(t_1)- \epsilon_{i\to i+1}(t_1) \right] .
		\label{Qdot1}
	\end{equation}
	Using eq. (\ref{full energy}) and the functional form of Green's functions in eq. (\ref{functional gf}) up to leading order in $1/q$, we get
	\begin{equation}
		\begin{aligned}
			\epsilon_{i-1\to i}(t_1)&= \int_{-\infty}^{t_1} dt_2 \hspace{1mm} 2\iu q D_{i-1}^\star(t_1) D_{i-1}(t_2) \left[  \Qq_{i-1}(t) - \Qq_i(t) \right] \\
			\epsilon_{i\to i+1}(t_1) &= \int_{-\infty}^{t_1} dt_2 \hspace{1mm} 2\iu q D_{i}^\star(t_1) D_{i}(t_2) \left[ \Qq_{i}(t) - \Qq_{i+1}(t) \right] 
		\end{aligned}
	\end{equation} 
	where we have previously defined $t \equiv (t_1 + t_2)/2$.
	By inserting this into eq. \eqref{Qdot1}, we obtain an explicit differential equation describing the change in local charge density for the leading order in $1/q$ (recall $r=\Oo(q^0)$)
	\begin{equation}
		\dot{\bm{\Qq}}(t_1) =	\frac{r}{q} \int_{-\infty}^{t_1}dt_2 \hspace{1mm} \left[ \bm{H}(t_1,t_2)\bm{\Qq}(t)+\Oo(q^{-1}) \right] 
		\label{charge dot}
	\end{equation}
	where
	\begin{widetext}
		\begin{equation}
			H_{ij}(t_1,t_2) = 2\Re [D_{i-1}(t_1) D_{i-1}^*(t_2)\delta_{j,i-1}+D_i(t_1) D_{i}^*(t_2)\delta_{j,i+1} -  \left( D_{i-1}(t_1) D_{i-1}^*(t_2)+D_i(t_1) D_{i}^*(t_2)\right) \delta_{j,i}]
		\end{equation}
	\end{widetext}
	where we note that $q \dot{\Qq}(t) = \Oo(\Qq)$. Note, that the above equation is closed under the local charge densities. In other words, the Green's functions do not enter into the expression to leading order in $1/q$. This is a drastic simplification of the general problem. The above equation implies that the local charge density can change on time scales $t = \Oo(q^0)$, but the fluctuations would then be of the order $\Oo(\Qq q^{-1})$. Hence, to leading order in $1/q$, the local charge density effectively remains constant \footnote{The physical reason for such a fine-tuned large-$q$ model construction can be expressed as essentially being a question about which terms will compete with one another. To have a competition between transport terms and onsite interactions on the level of the Green's functions, one must consider ``small'' transport terms. This leads to small charge fluctuations, but still keeps the influence of the transport on the level of the Green's functions. To have competition on both levels, one would have to consider $r$ scaling in $q$, something we plan to study in the future.}. In other words, if one does \emph{not} consider a rescaled time, $t \neq q^{3/2} \tau$, then for any finite time $t = \Oo(q^0)$, there is no charge flow. 
	
	Having obtained this general equation of motion, let us solve it for a particular case, which we will encounter again at a later stage. That is the case of a quench where we switch on the transport interactions at $t=0$
	\begin{equation}
		D_i(t) = R_i \Theta(t) \qquad (\text{Quench at } t=0)
		\label{quench}
	\end{equation}
	where $R_i$ are any arbitrary real or complex constants. Then taking the second derivative of eq. (\ref{charge dot}), we obtain the result for the charge transport dynamics as
	\begin{equation}
		\ddot{\bm{\Qq}} = \frac{r}{q} \bm{H} \bm{\Qq}
		\label{charge ddot}
	\end{equation}
	where
	\begin{equation}
		H_{ij} = |R_i|^2 \delta_{j,i+1}+|R_{i-1}|^2 \delta_{j,i-1}- \left[ |R_i|^2+|R_{i-1}|^2 \right] \delta_{ij}.
	\end{equation}
	We have thus obtained a \emph{discrete wave equation} independent of the on-site interaction strengths and depending only on the local charge densities as well as the transport coupling strengths. We have taken $R_0 = R_{2L+1} =0$. 
	
	To see explicitly how charge flows, let us consider $R_i = R$ $\forall$ $i$.  Then after the quench (eq. \eqref{quench}), the solution of eq. \eqref{charge ddot} to leading order in $1/q$ is given by
	\begin{equation}
		\Qq_i(t) = \Qq_i(0)+\frac{\Qq_{i+1}(0)+\Qq_{i-1}(0)-2\Qq_i(0)}{q}2 R^2 t^2.
	\end{equation}  
	as shown in App.\ref{appendix B}. Note that the only stationary state would correspond to a uniformly charged chain. In general, every site will gain charge assuming the neighboring sites have a combined larger charge density or otherwise lose charge.

	\section{A Note on Instantaneous Thermalization}
	\label{4}
	\subsection{Leading Order Kadanoff-Baym Equations}
	\label{4.1}
	We already have the functional form of the Green's functions (eqs (\ref{large q gf}, \ref{functional gf})). The purpose is to plug this in the KB equations (eq. (\ref{kb})) and derive a leading order in $1/q$ behavior for the KB equations. We reproduce the KB equations here as follows for convenience:
	
	\begin{equation}
		\begin{aligned}
			\left[\partial_{t_1} - \iu \dot{\eta}_i(t_1) \right]\Gg_i^{\gtrless}(t_1,t_2)
			=& \int_{t_1}^{t_2} dt_3 \hspace{1mm} \Sigma_i^{\gtrless}(t_1,t_3) \Gg_i^{A}(t_3,t_2) \\ &- \frac{\iu}{2 q} \alpha_i(t_1,t_2).
		\end{aligned}
	\end{equation}
	We start by evaluating the left-hand side of the KB equations using the functional form of Green's functions from eq. (\ref{large q gf}) where we find that the $\eta_i$ term cancels out to get
	$$\dot{\Qq}_{i}\left(t\right) e^{ \i\eta_{i}(t_1,t_2) + g_{i}^\gtrless(t_1,t_2)/q} + q^{-1} \Gg^{\gtrless}_{i}(t_1,t_2) \p_{t_1}g_{i}^\gtrless(t_1,t_2).
	$$
	Plugging back in the KB equations and rearranging yields
	\begin{widetext}
		\begin{equation}
			\p_{t_1}g^{\gtrless}_{i}(t_1,t_2) = \int_{t_1}^{t_2} dt_3 \hspace{1mm}\frac{q\Sigma_{i}^{\gtrless}(t_1,t_3)\Gg_i^{A}(t_3,t_2)}{\Gg^{\gtrless}_{i}(t_1,t_2)} - \left(\frac{\iu \alpha_i(t_1,t_2)/2+q\dot{\Qq}_{i}(t) e^{\i\eta_{i}(t_1,t_2) + g_{i}^\gtrless(t_1,t_2)/q}  }{\Gg^{\gtrless}_{i}(t_1,t_2)} \right).
			\label{KBq}
		\end{equation}
	\end{widetext}
	Up until this point, everything is exact. Now we start truncating at the leading order in $1/q$. We start by considering the functional form of Green's functions that appears in the denominator above, which at leading order is given by $\Gg_{i}^{\gtrless} \sim \mp 1/2$. We further recall that $\Qq = \Oo(q^{-1/2})$ while from the charge transport dynamical equation eq. (\ref{charge dot}), we have $q\dot{\Qq}_{i}(t) = \Oo(\Qq)$ $\Rightarrow \dot{\Qq}(t) = \Oo(q^{-3/2}) $. We also have $\eta_i = \Oo(q^{-1/2})$. Moreover, from the definition of $\alpha(t_1, t_2)$ in eq. (\ref{alpha}), we see that up to leading order, $\Gg_{i}^{\gtrless} \sim \mp 1/2$ there too (recall $\Sigma_i = \Oo(1/q)$), thereby making $\alpha(t_1, t_2)$ lose its $t_2$ dependence. Finally, using the definition of $\Gg_i^A(t_1, t_2)$ from eq. (\ref{advanced g}), $\Gg^{A}_{i}(t_3,t_2) = \Theta(t_2-t_3)$ at leading order in $1/q$. Thus the KB equations at leading order in $1/q$ become
	\begin{equation}
		\partial_{t_1}\frac{g^{\gtrless}_{i}(t_1,t_2)}{2} =  \mp \left( \int_{t_1}^{t_2} dt_3 \hspace{1mm}  q\Sigma_{i}^{\gtrless}(t_1,t_3) \right) \pm \frac{\iu \alpha_i(t_1)}{2} .
		\label{kb leading q}
	\end{equation}
	We can also express the self-energies $\Sigma^{\gtrless}_i(t_1, t_2)$ appearing here at the leading order using the explicit forms in eq. (\ref{se gtrless}). We can write $\Sigma^{\gtrless}_{i}(t_1, t_2) = \Sigma^{\gtrless}_{J,i}(t_1, t_2) + \Sigma^{\gtrless}_{D,i}(t_1, t_2)$ (eq. (\ref{se1})) where the on-site (same as eq. (\ref{functional se})) and transport contributions at leading order in $1/q$ are as follows:
	\begin{equation}
		\begin{aligned}
			q\Sigma_{D,i}^{\gtrless}(t_1,t_2) &= \mp \Dd_{\text{eff,}i}^2(t_1,t_2) \\
			q \Sigma_{J,i}^{\gtrless}(t_1,t_2) & = \mp \Jj^2_{\text{eff,}i}(t_1,t_2) e^{g^+_{i}(t_1,t_2)}
		\end{aligned}
		\label{se effective}
	\end{equation}
	respectively. Here the effective on-site and transport coupling strengths are
	
	\begin{equation}
		\begin{aligned}
			\Dd_{\text{eff,}i}^2(t_1,t_2) &\equiv D_{i-1}^*(t_1)D_{i-1}(t_2)+ D_{i}(t_1)D_{i}^*(t_2) \\
			\Jj^2_{\text{eff,}i}(t_1,t_2) &\equiv J_{i}(t_1)J_{i}(t_2)e^{-2q \Qq^2}
		\end{aligned}
	\end{equation}
	respectively.
	
	\subsection{Symmetric and Asymmetric Green's Functions}
	\label{4.2}
	We now introduce (a)symmetric Green's functions
	\begin{equation}
		g_i^{\pm}(t_1,t_2) \equiv \frac{g_i^>(t_1,t_2) \pm g_i^<(t_2,t_1)}{2} \label{g+-}
	\end{equation}
	which can be inverted to get
	\begin{equation}
		\begin{aligned}
			g_i^>(t_1,t_2) &=	g_i^{+}(t_1,t_2) + g_i^{-}(t_1,t_2) \\
			g_i^<(t_2,t_1) &=	g_i^{+}(t_1,t_2) - g_i^{-}(t_1,t_2).
		\end{aligned}
		\label{inverted}
	\end{equation}
	The physical motivation for introducing them comes from the fact that the Green's functions for Majorana fermions are symmetric but they are not symmetric for complex fermions which we are considering. Since Majorana Green's functions are symmetric, accordingly the asymmetric Green's function defined above vanishes for the Majorana case, namely $g_i^- =0$ \cite{Eberlein2017}. For complex fermions, $g_i^- \neq 0$ and the interpretation is that $g_i^-$ is in a sense a measure of deviations away from charge neutrality.
	
	We express eq. (\ref{kb leading q}) in terms of these new Green's functions by taking the derivative. Note that the order of taking derivative matters \footnote{For instance, if the derivative is to be taken with respect to, say $t_2$, which is the second argument in the KB equations, we first need to swap $t_1$ and $t_2$ in the arguments to keep $t_2$ at the first place to take the derivative and then take the conjugate.}. Recall $\mathcal{G}_i\left(t_1, t_2\right)^*=\mathcal{G}_i\left(t_2, t_1\right)$ and consequently $g^{\gtrless}_i\left(t_1, t_2\right)^*=g^{\gtrless}_i\left(t_2, t_1\right)$ \footnote{As shown in \cite{Eberlein2017, Louw2022}, $g^{>}_i\left(t_1, t_2\right)=g^{<}_i\left(t_2, t_1\right)$ holds true for Majorana fermions which can be derived from the complex SYK case by taking the limit $\Qq \rightarrow 0$.}. By doing this, we get
	
	\begin{widetext}
		\begin{equation}
			\begin{aligned}
				\partial_{t_1} g^{+}_{i}(t_1,t_2) = &   -\int_{t_1}^{t_2} dt_3 \hspace{1mm} \left( q\Sigma_{i}^{>}(t_1,t_3)-q\Sigma_{i}^{<}(t_3,t_1) \right)  +\iu \Re \left[ \alpha_i(t_1) \right] \\
				\partial_{t_1}g^{-}_{i}(t_1,t_2) = &  -\int_{t_1}^{t_2} dt_3 \hspace{1mm} \left( q\Sigma_{i}^{>}(t_1,t_3)+q\Sigma_{i}^{<}(t_3,t_1) \right) - \Im \left[ \alpha_i(t_1)  \right] .
			\end{aligned}
			\label{symmetric g}
		\end{equation}
	\end{widetext}
	
	Then we can also evaluate the equal-time KB equations when $t_2 \rightarrow t_1$ at leading order in $1/q$. Eq. (\ref{symmetric g}) reduces to
	\begin{align}
		\partial_t g_i^+(t_1, t_1^+) &= \iu \Re \left[ \alpha_i(t_1)\right] \label{equal time symmetric g}\\
		\partial_t g_i^-(t_1, t_1^+) &= - \Im \left[ \alpha_i(t_1)  \right] \label{dotg-}
	\end{align}
	where we take the limit $t_1^+ \rightarrow t_1$ only after taking the derivative. We already know the equal-time KB equations from eq. (\ref{kb energy}) where on-site and transport energies are given in eq. (\ref{full energy}). At leading order in $1/q$, we then get
	\begin{equation}
		\alpha_i(t_1,t_1) = \epsilon_i(t_1)+\frac{r}{q} \left[ \epsilon_{i\to i+1}^*(t_1)+\epsilon_{i-1\to i}(t_1) \right]
		\label{kb instant}
	\end{equation}
	where
	\begin{equation}
		\begin{aligned}
			\epsilon_i(t_1)  &= \Im\int_{-\infty}^{t_1}dt_2 \hspace{1mm} 2 \Jj^2_{\text{eff,}i} e^{g_i^+(t_1, t_2)}\\
			\epsilon_{i \rightarrow i+1}(t_1) &= 2 \left( \Qq_{i+1}(t) - \Qq_i(t) \right) \int_{-\infty}^{t_1} \hspace{-3mm} d t_2 \hspace{1mm} \iu q D_{i}^*(t_1) D_{i}(t_2)  .
		\end{aligned}
		\label{transport energy q}
	\end{equation}
	
	From the above expressions, eq. \eqref{transport energy q}, we note that the boundary condition on $g_i^+$ takes the form
	\begin{equation}
		\partial_t g_i^+(t_1, t_1^+) = \iu \epsilon_i(t_1) + \Oo(q^{-1}), \label{leadBoundCond}
	\end{equation}
	due to $\epsilon_{i \rightarrow i+1}(t_1)$ being imaginary to leading order.
	
	Finally, we take the second derivative of eq. (\ref{symmetric g}) with respect to $t_2$, where we use eq. (\ref{se effective}) for self-energies and the fact that complex conjugate amounts to switching of the two time arguments, to get (recall that partial derivatives commute)
	\begin{equation}
		\begin{aligned}
			\partial_{t_1}\p_{t_2}g^{+}_{i}(t_1,t_2)  & = 2\Re\left[ \Dd_i(t_1,t_2) + \Jj^2_{\text{eff,}i}(t_1,t_2) e^{g^+_{i}(t_1,t_2)} \right] \\
			\partial_{t_1} \partial_{t_2}g^{-}_{i}(t_1,t_2) &= 2 \iu \Im  \left[ \Dd_{\text{eff,}i}(t_1,t_2) \right].
			\label{secDer}
		\end{aligned}
	\end{equation}
	Let us recall the equation of motion describing the local change in charge density $\Im \left[ \alpha_i(t_1,t_1) \right] = q\dot{\Qq}_{i}(t_1)$ (eq. \eqref{Qdot}). Since this was of order $\Oo(\Qq)$, we note that the same order appears in the equal time derivative \eqref{dotg-}. In fact, it is known that for a single disconnected dot, to leading order in $1/q$, $g_i^- = \Oo (\Qq)$ \cite{Louw2022}. From \eqref{secDer}, we note that this result only extends over to the chain given real effective transport interactions $\Dd_{\text{eff,}i}^2(t_1,t_2) \in \RR$. Since we are restricting our analysis to the non-trivial on-site interactions where the local charge density scales as $\Qq = \Oo(q^{-1/2})$, assuming $\Dd_{\text{eff,}i}^2(t_1,t_2) \in \RR$, we have to leading order (using eq. (\ref{inverted}))
	\begin{equation}
		g_i^>(t_1,t_2)  \sim g_i^{+}(t_1,t_2), \quad g_i^<(t_2,t_1) \sim  g_i^{+}(t_1,t_2). \label{leadOrdsym}
	\end{equation}
	\phantom{ .}

	\subsection{Lack of Instantaneous Thermalization}
	\label{4.3}
	
	If $D_i=0$ $\forall$ $i$, then we are left with individual disconnected SYK blobs. We already know that a large-$q$ complex SYK model (which exists in each blob) instantaneously thermalizes \cite{Louw2022}. A natural question to ask is what happens in the case of our chain where we connect the large-$q$ SYK blobs with $r/2$-particle nearest neighbor hopping. We show via proof by contradiction that this is not the case here. 
	
	One might argue that since there exist charge fluctuations, that the system would clearly not be in thermal equilibrium. However, here we are considering the large-$q$ case, where the Green's functions change over a timescale $t=\Oo(q^0)$, while the fluctuations in charge density are of the order $\Oo(\Qq/q)$ at such time. As such, in this limit, the local charge densities are effectively constant, while the Green's functions are not. 
	
	For our proof, we again start with the same quench protocol as in eq. (\ref{quench}): $D_i(t) = R_i \Theta(t)$ where $R_i$ are any arbitrary real or complex constants. Therefore for $t<0$, we have disconnected large-$q$ SYK blobs which thermalize instantaneously thereby causing the whole system to be in equilibrium in pre-quench. Then at $t=0$, we connect them with $r/2$-particle nearest neighbor hopping term that leads to non-equilibrium dynamics in the system. 
	
	We focus on the real part of the transport energies, which is given by eq. (\ref{full energy})
	\begin{widetext}
		\begin{equation}
			\Re\left[ \epsilon_{i\to i+1}(t_1) \right]  = \int_0^{t_1} dt_2 \hspace{1mm} |R_{i}|^2 \left(\frac{-2q}{r} \right) \Im \underbrace{\left[ [-4\Gg_{i+1}^>(t_1,t_2)\Gg_{i}^<(t_2,t_1)]^{r/2}-[-4\Gg_{i}^>(t_1,t_2)^{\star} \Gg_{i+1}^<(t_2,t_1)^{\star}]^{r/2} \right]}_{r\boldsymbol{A}}.
			\label{contradiction}
		\end{equation}	
	\end{widetext}
	If we plug in the functional form of Green's functions from eq. (\ref{functional gf}), we find that the quadratic in charge density (recall $\Qq = \Oo(q^{-1/2})$) terms $\Qq_{i+1}^2$, $\Qq_i^2$ cancel, leaving 
	
	\begin{widetext}
		$$\bm{A} = 2[\Qq_{i+1}(t) - \Qq_i(t)]+ \frac{g_{i+1}^>(t_1, t_2)+g_i^<(t_2, t_1)}{2q}
		- \frac{g_{i}^{>}(t_1,t_2)^{\star}+g_{i+1}^{<}(t_2, t_1)^{\star}}{2q}.$$
	\end{widetext}

	The first term is the leading order contribution, which consists of local charge densities that are real. Since we are interested in the imaginary part of $\boldsymbol{A}$, this term drops out. The second and third terms are the next leading order contributions of $\boldsymbol{A}$ where clearly the $\Qq^2$ terms cancel out. Recall $t \equiv (t_1 + t_2)/2$. Thus, we are left with the $g^{\gtrless}$ terms in $\boldsymbol{A}$, namely
	\begin{equation}
		\frac{1}{2q} \left[  g_{i+1}^>(t_1, t_2) + g_i^<(t_2, t_1) - g_{i}^{>}(t_1, t_2)^{\star} - g_{i+1}^{<}(t_2, t_1)^{\star} \right].
	\end{equation}
	Note that such a quench to constant couplings yields real effective transport couplings $\Dd_{\text{eff,}i}^2(t_1,t_2) = |R_i|^2 + |R_{i+1}|^2 $. As such, the leading order equation for $g^-_i$ is of order $\Oo(\Qq)$. Since we are restricting our analysis to the non-trivial on-site interactions where the local charge density scales as $\Qq = \Oo(q^{-1/2})$, hence eq. \eqref{leadOrdsym} applies which states that all $g_i^{\gtrless}$ to leading order are given by their symmetric contributions leaving
	\begin{equation}
		\frac{1}{2q} \left[  g_{i+1}^+(t_1, t_2) + g_i^+(t_1, t_2) - g_{i}^{+}(t_1, t_2)^{\star} - g_{i+1}^{+}(t_1, t_2)^{\star} \right].
	\end{equation}
	This simplifies to
	\begin{equation}
		\frac{\iu }{q}\Im\left[ g_{i+1}^+(t_1,t_2)+g_{i}^+(t_1,t_2) \right] .
	\end{equation}
	Therefore plugging this back in eq. (\ref{contradiction}), we get
	\begin{equation}
		\begin{aligned}
			\Re & \left[   \epsilon_{i\to i+1} (t_1) \right]  \\
			&= \int_0^{t_1} dt_2 \hspace{1mm} |R_{i}|^2\Im \left[ g_{i+1}^+(t_1,t_2)+g_{i}^+(t_1, t_2) \right].
		\end{aligned}
		\label{e i i+1}
	\end{equation}
	In order to prove by contradiction, we now assume that the Green's function instantaneously thermalizes which implies that they can only depend on time differences, namely $g_i^+(t_1,t_2) = g_i^+(t_1-t_2)$ $\forall$ $i$. Then the real part of the local transport energy term becomes
	\begin{equation}
		\begin{aligned}
			\Re & \left[  \epsilon_{i\to i+1}  (t_1) \right]  \\
			&= \int_0^{t_1} dt_2 \hspace{1mm} |R_{i}|^2\Im \left[ g_{i+1}^+(t_1-t_2)+g_{i}^+(t_1 - t_2) \right] \\
			&= \int_0^{t_1}d\tau \hspace{1mm} |R_{i}|^2\Im \left[ g_{i+1}^+(\tau)+g_{i}^+(\tau) \right] .
		\end{aligned}
	\end{equation}
	But since we have assumed instantaneous thermalization, we know that the time derivative of the real part of the transport energy should be equal to zero. So we proceed to calculate the derivative of $\Re\left[ \epsilon_{i\to i+1}  (t_1) \right] $ with respect to $t_1$
	\begin{equation}
		\Re\left[ \dot{\epsilon}_{i\to i+1}  (t_1) \right] = |R_i|^2 \Im \left[ g_{i+1}^+(t_1)+g_{i}^+(t_1) \right] 
	\end{equation}
	But we saw in eq. (\ref{equal time symmetric g}) that the time derivative of $g^+_i$ is given as: $\partial_t g_i^+(t, t^+) = \iu \Re \left[ \alpha_i(t)\right]$. At $t=0^+$, there are only on-site interactions, therefore this reduces to (recall the discussion below eq. (\ref{inverted}) that $g_i^+ = \Oo(q^0)$)
	\begin{equation}
		\dot{g}_i^+(0^+) = \iu \Re\left[ \epsilon_i(0^+)\right] + \Oo(q^{-1}).
	\end{equation}
	as seen from \eqref{leadBoundCond}. Finally, taking another derivative of the transport energy, we get
	\begin{equation}
		\Re \left[ \ddot{\epsilon}_{i\to i+1}(0^+) \right]  = |R_{i}|^2\Im \left[ \dot{g}_{i+1}^+(0^+)+\dot{g}_{i}^+(0^+) \right]
	\end{equation}
	thereby leading to
	\begin{equation}
		\Re \left[ \ddot{\epsilon}_{i\to i+1}(0^+) \right]  = |R_{i}|^2 \left( \Re \left[ \epsilon_{i+1}(0^+) \right] +\Re \left[\epsilon_i(0^+) \right] \right).
		\label{final condition}
	\end{equation}
	But as aforementioned, $\Re \left[ \ddot{\epsilon}_{i\to i+1}(0^+) \right]  = 0$ if the system thermalizes instantaneously after the quench at $t=0^+$. This implies that if $R_i \neq 0$, then $\epsilon_{i+1}(0^+)$ and $\epsilon_i(0^+)$ must have opposite signs. However, for any positive temperature, the energy densities are always negative. Hence the assumption that the system thermalizes instantaneously is false and there is a lack of instantaneous thermalization for the chain even though individual blobs thermalizes instantaneously in isolation. This is also captured by the observation that $\Re \left[ \ddot{\epsilon}_{i\to i+1}(0^+) \right]$ vanishes when $R_i=0$ which simply reproduces the result for an individual large-$q$ SYK model as expected \cite{Louw2022}. 
	
	Another interesting possibility is that $J_i=0$ $\forall$ $i$ which will also satisfy the condition in eq. (\ref{final condition}). The interpretation of this result is that in the case of a pure transport chain of $r/2$ particles to nearest neighbors, \emph{can} lead to instantaneous thermalization. But having obtained the result for local charge density in eq. (\ref{charge ddot}), we know that there is a flow of current for any \emph{finite} $q$ that shows that the chain cannot be in equilibrium. But in the limit $q\rightarrow \infty$, the local charge density effectively becomes constant. Note that eq. (\ref{final condition}) does not rule out the possibility for a pure transport chain to thermalize instantaneously, in this limit.
	
	Lastly, note that at uniform coupling and charge density that our system effectively describes a single SYK dot. This can be seen in all the equations of motion reducing to that of a single dot together with kinetic-type and large $q$ coupling. As such, the proof by contradiction remains valid for a single dot. 
	
	In the general case, however, where we have both onsite and transport terms, the picture would be the following. While the total energy remains conserved, there exist fluctuations between the onsite and kinetic energies (eq. (\ref{ener}))
	\begin{equation}
		\epsilon_{i}(t_1) \propto   \langle \Hh_i\rangle(t_1), \quad \Re \left[ \epsilon_{i\to i+1}(t_1) \right] \propto \Re\langle \Hh_{i\to i+1} \rangle(t_1),
	\end{equation}
	which tend to their final values over a non-zero finite time. A natural question is how, for instance at what rate, these terms tend to their equilibrium values to attain thermalization. Such an analysis could be carried out by explicitly solving for the Green's functions. Alternatively, one may consider a linear stability analysis around the thermal Green's functions.
	
	\section{Generalizing the chain to a higher dimensional lattice}
	\label{general}
	
	Let us now consider the same model by on a $d$-dimensional lattice $\Lambda$ with the nearest neighbor hopping, where the Hamiltonian is given by
	\begin{equation}
		\Hh(t) = \sum_{x \in \Lambda} \Hh_{x}(t) + \sum_{\langle x,x'\rangle \in \Lambda}\Hh_{x \to x'}(t) 
		\label{HamChain2}
	\end{equation}
	where $\langle x,x'\rangle$ denotes nearest neighbor interactions. The explicit form of $\mathcal{H}_x$ is the same as in eq. (\ref{hi}) and the transport Hamiltonian from site $x$ to $x'$ is given by
	\begin{equation}
		\begin{aligned}
			\Hh_{x \rightarrow x'}(t) 
			&=  \hspace{-1mm} \sum\limits_{\substack{ \{\bm{\mu}\}_1^{r/2} \\ \{\bm{\nu}\}_1^{r/2} }} \hspace{-2mm}Y(x,x')^{\bm{\mu}}_{\bm{\nu}} c^{\dag}_{x';\mu_1} \cdots c^{\dag}_{x'; \mu_{\frac{r}{2}}} c_{x;\nu_{\frac{r}{2}}}^{\vphantom{\dag}} \cdots c_{x;\nu_1}^{\vphantom{\dag}},
		\end{aligned}
	\end{equation}
	Here $D_{xx'}(t)^* = D_{x'x}(t)$ and $(Y(x,x')^{\bm{\mu}}_{\bm{\nu}})^* = (Y(x',x)^{\bm{\nu}}_{\bm{\mu}})$ ensures hermiticity in the Hamiltonian. In the old notation, we always defined $D_{i}$ as the coupling corresponding to right hopping. To make it explicitly clear, for $d=1$ case, we had $D_{i-1}^*(t) = D_{i,i-1}(t)$ while $D_{i}(t) = D_{i,i+1}(t)$. All expressions remain unchanged, except that we now sum over $z = 2d$ nearest neighbors (which in the one-dimensional case would reduce to two nearest neighbors). So for instance, the action corresponding to site $x$ (eq. \eqref{noninteracting action} and \eqref{interacting action}) remains unchanged, only with $i$ being replaced by $x$. The transport Lagrangian $\Lll_{I;x}$ in \eqref{interacting action} however gains additional terms due to the now $z$ nearest neighbors
	\begin{equation}
		\begin{aligned}
			\sum_{x' :\langle x,x'\rangle} \frac{D_{xx'}(t_1)D_{xx'}^*(t_2)}{qr} [-4\Gg_{x'}(t_1,t_2) \Gg_{x}(t_2,t_1)]^{r/2}.
			\label{interacting lagrangian2}
		\end{aligned}
	\end{equation}
	Here $x' :\langle x,x'\rangle$ means summation is over $x'$ such that $x'$ is the nearest neighbor of $x$. This then yields the new transport self-energy for site $x$
	\begin{widetext}
		\begin{equation}
			\begin{aligned}
				q\Sigma_{D, x}(t_1,t_2) =  \sum_{x' :\langle x,x'\rangle} D_{xx'}(t_1)D_{xx'}^*(t_2) [-4\Gg_{x}(t_2,t_1) \Gg_{x'}(t_1,t_2)]^{\frac{r}{2}-1} 2\Gg_{x'}(t_1,t_2).
			\end{aligned}
			\label{se4}
		\end{equation}
	\end{widetext}
	As an example, for an equilibrium and translationally invariant system, $\Gg_{x}(t_1,t_2) = \Gg(t)$, we simply have $q\Sigma_{D, x}(t_1,t_2) =  z |D_{12}|^2 [-4\Gg(-t) \Gg(t)]^{\frac{r}{2}-1} 2\Gg(t)$.
	
	With this setup for a general nearest neighbor $d$-dimensional lattice $\Lambda$, we carry out the same analysis for the charge dynamics as in Sec. \ref{3}. We find that the analysis goes through, and we still are left with a closed-form equation for charge transport to leading order in $1/q$ as in eq. (\ref{charge dot}). Explicitly we have for site $x$
	\begin{equation}
		\dot{\Qq}_x(t_1) =	\frac{r}{q} \int_{-\infty}^{t_1}dt_2 \hspace{1mm} \sum\limits_{y \in \Lambda} \left[ H_{xy}(t_1,t_2)\Qq_{y}(t)+\Oo(q^{-1}) \right] 
		\label{charge dot lattice}
	\end{equation}
	where
	\begin{equation}
		H_{xy}(t_1,t_2) =  \sum\limits_{x':\langle x,x' \rangle}2\Re [D_{xx'}^*(t_1)D_{xx'}(t_2)] (\delta_{x^{\prime}y} - \delta_{xy})
	\end{equation}
	Thus we see that generalizing to a higher dimensional lattice preserves the closed-form relation for charge dynamics.
	
	Moreover, the result for lack of instantaneous thermalization as done in Sec. \ref{4} still holds for such a higher dimensional lattice $\Lambda$. We start with the equal-time KB equation at as in eq. (\ref{kb energy}) for site $x$ which is given by
	\begin{equation}
		\alpha_x(t_1,t_1) = \epsilon_x(t_1)+\frac{r}{q} \sum\limits_{j = 1}^d  \left( \epsilon_{x  \to x +  \hat{e}_j}^*(t_1)+ \epsilon_{x - \hat{e}_j \to x}(t_1) \right)
	\end{equation}
	where $\hat{e}_j$ is the unit vector pointing towards the positive direction along the dimension $j$ and the sum is over all possible dimensions. Then proceeding in the same manner as in Sec. \ref{4} for the same quench considered there (eq. \eqref{quench}), we note that eq. \eqref{e i i+1} is symmetric under the operation $i \longleftrightarrow i+1$. Therefore we get a similar equation as in eq. (\ref{e i i+1}) for site $x$ and some neighboring site $x'$
	\begin{equation}
		\begin{aligned}
			\Re  & \left[  \epsilon_{x\to x'} (t_1) \right]  \\
			&= \int_0^{t_1} dt_2 \hspace{1mm} |R_{xx'}|^2\Im \left[ g_{x'}^+(t_1,t_2)+g_{x}^+(t_1, t_2) \right].
		\end{aligned}
	\end{equation}
	
	Then again to prove by contradiction, we assume instantaneous thermalization after the aforementioned quench so that $g_x(t_1, t_2) = g_x(t_1-t_2)$ $\forall$ $x \in \Lambda$. We again get
	\begin{equation}
		\Re \left[ \ddot{\epsilon}_{x\to x'}(0^+) \right]  = |R_{xx'}|^2 \left( \Re \left[ \epsilon_{x'}(0^+)+\epsilon_x(0^+) \right] \right).
	\end{equation}
	But due to the assumption of instantaneous thermalization after the quench at $t=0$, we must have $\Re \left[ \ddot{\epsilon}_{x\to x'}(0^+) \right] = 0$ and this implies that if $R_{xx'} \neq 0$, then $ \epsilon_{x}(0^+) $ and $ \epsilon_{x'}(0^+) $ must have opposite signs. However, the on-site energy for any site $x \in \Lambda$ is always negative for any positive temperature. Thus our proof for lack of instantaneous thermalization holds true for any higher dimensional nearest neighbor lattice consisting of large-$q$ complex SYK models.

	\section{Conclusion and Outlook}
	\label{5}
	We considered in this work a chain of large-$q$ SYK dots connected by $r/2$-particles hopping to nearest neighbors. We assumed that $r$ does not scale with $q$ so that $r = \Oo(q^0)$. We already know that the case of $r=2$ amounts to quadratic hopping, which has been shown to exhibit strange metal behaviors \cite{Song2017}. We considered an even more general case of $r/2$-particles hopping, where we developed a rather general analytical framework and obtained the dynamical results at leading order in $1/q$. Surprisingly, we found that the physics of the diffusive $r>2$ is effectively the same as the kinetic case $r=2$ at leading order in $1/q$, assuming $r=\Oo(q^0)$.
	
	In Sec. \ref{2}, we have developed a rather general framework of dealing with a general chain as described above. Starting with calculating the effective action in the large-$\Nn$ limit, we calculated the Schwinger-Dyson equations that translated to the Kadanoff-Baym (KB) equations using the Langreth rules. Providing explicit expressions for self-energies, we evaluated the expectation values of energies and showed their connection with the equal-time KB equations. Working with a functional form for Green's functions, we were able to study the KB equations in the large-$q$ limit in Sec. \ref{4.1} that controls the non-equilibrium dynamics of the system at the leading order in $1/q$. Dealing with complex fermions necessitated the introduction of symmetric and antisymmetric Green's functions $g_i^{\pm}$ in Sec. \ref{4.2} where we evaluated their equations of motion. We gave an interpretation for both $g_i^{\pm}$ and expressed the leading order KB equations in terms of $g_i^{\pm}$.
	
	With this rather general framework developed, we proceeded to study the quench dynamics of the system. The quench is given in eq. (\ref{quench}) where we have instantaneously thermalized and disconnected large-$q$ complex SYK blobs for $t<0$. Then the transport coupling of $r/2$ particles to nearest neighbors is switched on at $t=0$. We found closed-form expressions for the local charge transport dynamics in a general scenario in eq. (\ref{charge dot}) and consequently eq. (\ref{charge ddot}) for quench dynamics. The corresponding closed-form result obtained in eq. (\ref{charge ddot}) is quite fascinating in the sense that this is a discrete wave equation that is completely independent of the on-site couplings $J_i$. So the $r/2$-particles charge transport somehow does not feel the on-site coupling strengths of the individual SYK blobs. Furthermore, we see that for any finite $q$, there is indeed a local change in charge density, albeit of the order $\Oo(\Qq/q)$. Only in the limit $q \rightarrow \infty$, do these fluctuations become vanishing small.
	
	Having known that a single large-$q$ SYK model instantaneously thermalizes \cite{Louw2022}, we asked the natural question about our chain in consideration. We again considered the same aforementioned quench and \emph{assumed} that the system does indeed thermalize instantaneously. This led us to the consistency relation (\ref{final condition}) which must vanish for the system to be in equilibrium. Thus we realized that if $R_i \neq 0$ and $J_i \neq 0$, then the on-site energies must have opposite signs to cancel each other but this cannot be true because, for any positive temperature, the on-site energy densities are always negative. Hence by contradiction, we proved that the chain does not instantaneously thermalize. The consistency condition (eq. (\ref{final condition})) also provides the necessary condition when it is satisfied by either $R_i = 0$ or $J_i = 0$. First considering the case of $R_i=0$, this means that the system has disconnected SYK blobs which we already know from \cite{Louw2022} that individually all the blobs instantaneously thermalize. The other  $J_i=0$ implies that the system is a pure transport chain of $r/2$ particles hopping to nearest neighbors. But having obtained the result for local charge density in eq. (\ref{charge ddot}), we know that there is a flow of current for any \emph{finite} $q$ that shows that the chain cannot be in equilibrium. But in the limit $q\rightarrow \infty$, the local charge density effectively becomes constant. Our result obtained in eq. (\ref{final condition}) does not rule out the possibility for a pure transport chain to thermalize instantaneously, in this limit.
	
	Finally, in Sec. \ref{general}, we generalized our analytical framework from a one-dimensional nearest neighbor chain to an arbitrary $d$-dimensional nearest neighbor lattice $\Lambda$. To leading order in $1/q$, we found that the equations describing the local charge density still remain closed, as explicitly shown in eq. (\ref{charge dot lattice}). Moreover, our proof by contradiction for the chain to show a lack of instantaneous thermalization, after a quench, also holds true for the lattice $\Lambda$. 
	
	We have solved for the case of $r/2$-particle hopping. This is the general \emph{diffusive} case where $r$ scales as $\Oo(q^0)$. The subset case of $r=2$ is the kinetic hopping, which has been studied in \cite{Song2017} to exhibit strange metal behaviors. One of the natural generalizations of this work is when $r$ scales as $q$ such that $r = \kappa q$ where $\kappa$ is some scalar constant. Moreover, we know that the SYK model shows maximally chaotic behavior \cite{Maldacena2016a}. We did not address the chaotic behavior of the chain that might lead to chaotic-integrable phase transitions as observed in a single SYK model \cite{Sorokhaibam2020}. We already know for a single large-$q$ complex SYK model that the critical exponents corresponding to this phase transition belong to the same universality class as that of AdS black holes \cite{Louw2023}, so a natural question to ask is what happens if we connect those SYK models in the form of a chain as considered in this work. Another crucial feature of the SYK model is that it serves as a model for quantum holography \cite{Kitaev2015}. This begs a natural question that whether the chain that we have considered does have a holographic dual or not. We leave these to future works.

	
	\begin{acknowledgments}
		This work was funded by the Deutsche Forschungsgemeinschaft (DFG, German Research Foundation) - SFB 1073. We are grateful to Stefan Kehrein for insightful discussions. 
	\end{acknowledgments}
	
	
	\appendix
	
	\section{Combinatorial Argument for Any SYK Type Effective Action}
	\label{appendix A}

	\subsection{Combinatorics of a disordered averaged action}
	\label{a.1}
	
	Here we describe the mathematics describing the leading order SYK action. We do this in a rather abstract combinatorial way and then proceed to relate it to the SYK case.
	
	Let us consider an action $\Aa = \sum_{i=1}^{\Nn} S_i$ where $S_i$ are identically distributed and independent random variables with zero mean. They scale as $\Nn^{-1/2}$ which ensures an extensive averaged action.  We would like to evaluate the following average:
	\begin{equation}
		e^{- S_{\text{eff}}} \equiv \overline{e^{\i \Aa}} = \sum_{n=0}^\infty \frac{\i^{2n}}{(2n)!}\overline{\Aa^{2n}} 
		\label{avAc}
	\end{equation}
	where we have already taken into account that odd powers average to zero due to the zero mean. We note that whenever two random variables average together, this implies that they are the same random variables with the same indices, reducing the number of free summations. For instance, a simple case will be
	\begin{equation}
		{\Aa^4} = \overline{\sum_{i_1,i_2,i_3,i_4} S_{i_1} S_{i_2} S_{i_3} S_{i_4}}
	\end{equation}
	There are $3$ ways in which two random variables average together and $1$ way where all of them average together, namely when $i_1 = i_2 = i_3 = i_4$. This implies that
	\begin{equation}
		{\Aa^4} = 3\sum_{i_1,i_2} \overline{S_{i_1}^2}\, \overline{S_{i_2}^2} + \sum_{i_1} \overline{S_{i_1}^4}
	\end{equation}
	Each sum contributes a factor $\Nn$, hence $\overline{S^4} = 3 \Nn^2 \overline{S^2}\, \overline{S^2} + \Nn \overline{S^4}$. Each power scales as $\overline{S^n} \sim \Nn^{-n/2}$, indicating that the first term is of order $\Oo(\Nn^{0})$ while the second term is over order $\Oo(1/\Nn)$ thereby becoming irrelevant for large $\Nn$. Hence in situations like these, we need only consider the averages between two random variables.  The same holds true in the SYK model where to leading order in $1/\Nn$, we only have to consider \emph{two} random variables averaging together (recall that the odd numbers of random variables average to zero). There are $(2n-1)(2n-3)\cdots 1$ different pairs of $S_i S_j$. Written differently, this implies that
	\begin{equation}
		{\Aa^{2n}} \sim \frac{(2n)!}{2^n n! } \overline{\Aa^{2}}^n
	\end{equation}
	hence reducing eq. (\ref{avAc}) to the effective action $S_{\text{eff}} = \overline{\Aa^2}/2$. Note that this argument does not rely on having Gaussian random variables. 
	
	\subsection{Relation with SYK Type Action}
	\label{a.2}
	
	The above problem relates to any SYK-type actions where the Hamiltonian is described by the Grassmann expectation value
	\begin{equation}
		(\bar{\psi}(t_1),\psi(t_1)) = \sum_{i=1}^{N} X_i F_i(\bar{\psi}(t_1),\psi(t_1))
	\end{equation}
	where $X_i$ is the random variable and $F_i$ is an arbitrary function of $\bar{\psi}$ and $\psi$. What this means is that for any SYK type Hamiltonian, the effective averaged action is given as follows:
	\begin{equation}
		S_{I} = \int dt_1 dt_2 \hspace{1mm} \frac{1}{2}\overline{\Hh(\bar{\psi}(t_1),\psi(t_1))\Hh(\bar{\psi}(t_2),\psi(t_2))}.
	\end{equation}
	
	\subsection{Lagrange Multipliers \& Effective Action}
	\label{a.3}
	
	Even when we combine many SYK models such as a chain as done in this work, the partition function takes on the form
	\begin{equation}
		\Zz = \int \mathcal{D}(\bar{\Psi},\Psi) e^{-S_I[G] + \text{Tr}\{\hat{\Gg}_0^{-1}\circ G\}}
	\end{equation}
	where the quadratic field $G$ is given by
	\begin{equation}
		G_{ij}(t_1,t_2) \equiv -\frac{1}{N} \sum\limits_{\alpha=1}^N \bar{\psi}_{i;\alpha}(t_1) \psi_{j;\alpha}(t_2)
		\label{gij}
	\end{equation}
	
	The matrix field multiplication is defined as
	\begin{align}
		(\hat{A}\circ \hat{B})(t_1,t_2) &= \int_{\Cc} dt_3 \hspace{1mm} \hat{A}(t_1,t_3) \hat{B}(t_3,t_2)
	\end{align} 
	which defines the trace over matrix fields' time components as
	\begin{equation}
		\text{Tr}\{\hat{A} \circ \hat{B}\} \equiv \int_{\Cc} dt_1 \hspace{1mm} (\hat{A}\circ \hat{B})(t_1,t_1)
	\end{equation}
	
	Lagrange multiplier $\Gg$ is introduced via a delta functional which further is expressed in terms of another Lagrange multiplier $\hat{\Sigma}$ as follows
	\begin{align*}
		\Zz &= \int \mathcal{D}(\bar{\Psi},\Psi) \int d\hat{\Gg} \delta[\hat{\Gg} - G] e^{\text{Tr}\{\hat{\Gg}_0^{-1}\circ \,G\}} e^{-S_I[\hat{\Gg}]}\\
		&= \int \mathcal{D}(\bar{\Psi},\Psi) \int d\hat{\Gg} \int d\hat{\Sigma} e^{\text{Tr}\{\hat{\Sigma} \circ [\hat{\Gg}-G]\}}  e^{\text{Tr}\{\hat{\Gg}_0^{-1}\circ G\}}e^{-S_I[\hat{\Gg}]}\\
		&= \int d\hat{\Gg} \int d\hat{\Sigma}  e^{-S_0[\hat{\Gg},\hat{\Sigma}]} e^{-S_I[\hat{\Gg}]}
	\end{align*}
	where we have Gaussian-type integrals of Grassmann fields in the definition of $G$ (eq. (\ref{gij})). After integrating out the Grassmann fields, we get the \emph{effective non-interacting action}
	\begin{equation}
		S_0[\Gg,\Sigma] \equiv -\text{Tr}\{\hat{\Sigma} \circ \hat{\Gg}+\ln[\hat{\Gg}^{-1}_0 - \hat{\Sigma}]\}    
	\end{equation}
	By varying the action w.r.t. $\hat{\Sigma}$, we obtain the \emph{Dyson's equation}
	\begin{equation}
		\hat{\Gg} -[\hat{\Gg}^{-1}_0 - \hat{\Sigma}]^{-1} = 0
	\end{equation}

	\section{Charge Transport Solution}
	\label{appendix B}
	
	Here we provide the full solution to the vector equation $\ddot{\bm{\Qq}} = \bm{H} \bm{\Qq}$, with
	\begin{equation} 
		H_{ij} = \frac{4}{q} \left[R_i^2 \delta_{j,i+1}+R_{i-1}^2 \delta_{j,i-1}-[R_i^2+R_{i-1}^2] \delta_{ij}\right] 
	\end{equation}
	where $R_0=R_{2L+1}=0$ and $R_i = R$ $\forall$ $i$. Then
	\begin{equation}
		H = -\frac{16 R^2}{q} M, \quad M = \frac{\mathds{1}}{2} - \frac{1}{4}\begin{bmatrix}1 & 1 & 0 & \ldots & 0 & 0\\ 1 & 0 & 1 & 0 & \ldots & 0\\ 0 & 1 & 0 & 1 & 0 & \ldots\\  &  &  & \ddots & &\\ 0 & \ldots & 0 & 1 & 0 & 1\\ 0 & \ldots &0 & 0 & 1 &1 \end{bmatrix}
	\end{equation}
	where charge conservation is seen in all columns summing to zero. The near Toeplitz matrix $M$ can be diagonalized $\Lambda = U M U^{\dag}$ via the unitary matrix $U_{ij}$. With this, the solution to the time-dependent charge density is given by
	\begin{equation}
		\bm{\Qq}(t) = U^\dag\cos (\tau \sqrt{\Lambda}) U\bm{\Qq}(0)\qquad \left( \tau \equiv \frac{4 R}{\sqrt{q}} t \right)
	\end{equation}
	or explicitly:
	\begin{align}
		\Qq_i(t) &= \sum_{j=1}^{2L}c_{ij}(\tau)\Qq_j(0)\\
		c_{ij}(\tau) &\equiv \sum_{k=1}^{2L}  \cos (\tau \sqrt{\Lambda_{kk}}) U^\dag_{ik} U_{kj}.
		\label{coeff}
	\end{align}  
	
	Now we need to find an explicit form of $U_{ij}$. We define  $U_{1j} = 1/\sqrt{2L}$ and for $k\neq 1$
	\begin{equation}
		U_{kj} = \sqrt{\frac{1}{L}} \cos\left(\frac{p_{k-1}}{2} [2j-1] \right), \quad p_k \equiv \frac{\pi k}{2L}.
	\end{equation}
	Then the diagonalized matrix $\Lambda$ is the matrix of eigenvalues which is given by
	\begin{equation}
		\Lambda_{kk} = \sin^2\left(p_{k-1}/2\right)
	\end{equation}
	The coefficients in eq. (\ref{coeff}) become
	\begin{equation}
		c_{ij}(\tau) \equiv \sum_{k=1}^{2L}  \cos (\tau \sqrt{\Lambda_{kk}}) U^\dag_{ik} U_{kj}
		=\frac{1}{L} + d_{ij}(\tau)
	\end{equation}
	with
	\begin{widetext}
		\begin{equation}
			d_{ij}(\tau) \equiv \frac{2}{\pi} \frac{\pi}{4 L} \sum_{m=0}^{2L-1} 2\cos\left[\tau \sin\left[\frac{p_{m}}{2}\right]\right]  \cos\left[\frac{p_{m}}{2} [2i-1] \right] \cos\left[\frac{p_{m}}{2} [2j-1] \right]
		\end{equation}
	\end{widetext}
	where the inner term may be expressed as
	\begin{equation}
		\begin{aligned}
			2  \cos&\left[x [2i-1] \right] \cos\left[x [2j-1] \right]\\ =&  \cos[2(i-j)x] +  \cos[2 (i+j-1) x]
		\end{aligned}
	\end{equation}
	
	Now for large $2L$ the sum
	\begin{equation}
		\begin{aligned}
			\frac{2}{\pi} \frac{\pi}{4 L} \sum_{m=0}^{2L-1} \cos\left[\tau \sin (p_m/2)\right] \cos(2 n p_m/2)
		\end{aligned}
	\end{equation}
	can be approximated by the integral
	\begin{equation}
		\begin{aligned}
			A_{n}(\tau) &=\frac{2}{\pi} \int_0^{\pi/2}dx \hspace{1mm} \cos\left[\tau \sin x\right] \cos(2 n x)\\
			&= J_{2n}(\tau)
		\end{aligned}
	\end{equation}
	where the Bessel functions are defined by
	\begin{equation}
		J_{n}(\tau) \equiv \frac{1}{\pi} \int_0^{\pi}dx \hspace{1mm}\cos(n x - \tau \sin x)
	\end{equation}
	
	So we have that for large $2L$
	\begin{equation}
		c_{ij}(\tau) \sim J_{2|i-j|}(\tau) + J_{2|i+j-1|}(\tau)
	\end{equation}
	
	Total charge conservation is then ensured by the Bessel function property \cite{Abramowitz1965}
	\begin{equation}
		1 = J_0(\tau) + 2 \sum_{n=1}^\infty J_{2n}(\tau)
	\end{equation}
	from which one can show that
	\begin{equation}
		\Qq(t) = \sum_{j=1}^{2L} \sum_{i=1}^{2L} c_{ij}(\tau)\Qq_j(0), 
	\end{equation}  
	is equal to $\Qq(t)$, by showing that $ \sum_{i=1}^{\infty} c_{ij}(\tau) = 1$. To leading order we have $J_{n}(\tau) = (\tau/2)^{n}/n!+\Oo(\tau^{n+2})$. As such, if $\tau = \Oo(q^{-1/2})$, then the dynamics of $\Qq_i$ are dominated by
	\begin{equation}
		c_{ii}(\tau) \sim 1 - \tau^2/4 \quad c_{i,i\pm 1}(\tau) \sim \tau^2/8
	\end{equation}
	Explicitly we have
	\begin{equation}
		\Qq_i(t) = \Qq_i(0)+\left[\Qq_{i-1}(0)-2\Qq_i(0)+\Qq_{i+1}(0)\right]\tau^2/8
	\end{equation}

	\bibliography{refs.bib}

\end{document}